\documentstyle[12pt]{article}
\renewcommand
\baselinestretch{1.3}

\begin{document}

%\hfill UB-ECM-PF 93/1

\hfill HUPD-93-03

\hfill January 1993

\vspace*{3mm}

\begin{center}

{\LARGE \bf
Convenient versus unique effective action formalism in 2d
dilaton-Maxwell
quantum gravity}
\vspace{4mm}

\renewcommand
\baselinestretch{0.8}
{\sc E. Elizalde}\footnote{E-mail address: eli @ ebubecm1.bitnet}
\\
{\it Department E.C.M., Faculty of Physics, University of
Barcelona, \\
Diagonal 647, 08028 Barcelona, Spain} \\
{\sc S. Naftulin}
\\
{\it Institute for Single Crystals, 60 Lenin Ave., 310141 Kharkov,
Ukraine} \\  and \\
{\sc S.D. Odintsov}\footnote{On sabbatical leave from
Tomsk Pedagogical Institute, 634041 Tomsk, Russia. E-mail address:
odintsov @ theo.phys.sci.hiroshima-u.ac.jp} \\ {\it Department of
Physics, Faculty of Science, Hiroshima University, \\
Higashi-Hiroshima 724, Japan}
\medskip

\renewcommand
\baselinestretch{1.4}

\vspace{5mm}

{\bf Abstract}

\end{center}

The structure of one-loop divergences of two-dimensional
dilaton-Maxwell quantum
gravity is investigated in two formalisms:  one using a convenient
effective action and the other a unique effective action. The
one-loop divergences (including surface divergences)
of the convenient effective action are calculated in three
different covariant gauges: (i) De Witt, (ii) $\Omega$-degenerate
De Witt, and (iii) simplest covariant. The on-shell effective
action is given by surface divergences only (finiteness of the
$S$-matrix), which yet depend upon the gauge condition choice.
 Off-shell renormalizability is discussed and classes of
renormalizable dilaton and Maxwell potentials are found which
coincide in the cases of convenient and unique effective actions.
A detailed comparison of both situations, i.e. convenient vs.
unique effective action, is given. As an extension of the
procedure, the one-loop effective action in two-dimensional
dilaton-Yang-Mills gravity is calculated.

\vspace{3mm}

\noindent PACS: 04.50, 03.70, 11.17.

\newpage

\section{Introduction}

Black hole physics and early universe physics are issues in which
quantum gravity effects are expected to be significant. Some years
have elapsed since the earliest attempts [1-3] and we still do not
have a consistent description of quantum black holes (specially, of
their final states). One of the main reasons is that gravity must
be quantized for such considerations; however, Einstenian quantum
gravity is not consistent due to non-renormalizability in four
dimensions, and modifications of the same still have similar (or
sometimes additional) drawbacks.

In such situation it is quite advisable to study solvable toy
models where all these problems may be simplified drastically while
keeping still many of the good properties of the more realistic
theory.
Two-dimensional (2d) quantum gravity with matter can be considered
as such a toy model [4]. This theory is multiplicatively
renormalizable, and contains black holes and Hawking radiation [1].
Starting from the seminal work of Callan, Giddings, Harvey and
Strominger (CGHS) [4] where these properties were first realized,
some interesting additional results about the CGHS model and its
modifications [5-7] have been obtained (for a review, see [7]).
In particular, 2d black holes ---previously found in the string
context [8] (for an earlier discussion, see [31])--- and their
properties
 have been investigated
intensively.

In Refs. [9,11] black hole solutions have been
obtained in 2d dilaton-Maxwell gravity with the action
\begin{equation}
S=-\int d^{2}x \,\sqrt{g}\left[ {1\over 2} g^{\mu  \nu}\nabla_\mu
\Phi
        \nabla_\nu \Phi + CR\Phi +V(\Phi ) + {1\over 4}f(\Phi )
        g^{\mu\alpha}g^{\nu\beta}F_{\mu\nu}F_{\alpha\beta}\right].
%\label{}
\end{equation}
where $F_{\mu\nu}= \partial_{\mu} A_{\nu}-\partial_{\nu} A_{\mu}$,
$\Phi$ is the dilaton, and $V(\Phi)$ and $f(\Phi)$ are the
dilatonic and Maxwell potentials, respectively. Model (1) is
connected (via some compactification) with the four-dimensional
Einstein-Maxwell theory, which admits charged black hole solutions
[10]. Particular cases of (1) describe the bosonic string effective
action ($A_{\mu} =0$) and the heterotic string effective action
($A_{\mu} \neq 0$). The renormalization structure of 2d dilaton
gravity in covariant gauges has been investigated in Refs. [12-15].
It has been shown there that the theory is multiplicatively
renormalizable off-shell for the Liuoville dilaton potential (for
some specific choices of the parameters the theory is even finite
[14]). Moreover, the one-loop on-shell effective action is given by
surface counterterms only; that is why  the one-loop $S$-matrix is
finite, as in 4d Einstein gravity [18]. This shows explicitly that
2d quantum gravity can be a very good laboratory for studying the
formal structure of quantum field theory.

In the present paper we investigate the renormalization structure
of 2d dilaton-Maxwell gravity (1). In Sec. 2, we calculate the
one-loop divergences
 of this convenient effective action in two covariant
gauges (one of which is De Witt's gauge). All surface divergences
are taken into account in this consideration. (Note that similar
calculation in a simplest covariant gauge have been done in Refs.
[11,16]). These terms may be important in the discussion of the
Casimir effect [25]. In Sect. 3  we calculate the one-loop
divergences in the unique effective action formalism [19,20] (see
[21] for a review), which gives a parametrization and gauge
independent result. Finally, Sect. 4 is devoted to discussions and
to a comparison of the results obtained in the two formalisms. In
an Appendix we obtain the one-loop divergences of 2d Yang-Mills
dilaton gravity in the simplest covariant gauge.

\section{One-loop divergences of 2d dilaton-Maxwell gravity in the
convenient effective action formalism}

In this section we calculate the one-loop divergences (including
surface divergences) of the convenient effective action (1) for 2d
dilaton-Maxwell gravity in the covariant gauge. The classical
dynamics of the theory are defined by the following field
equations
\begin{eqnarray}
&& \nabla_{\mu} (f F^{\mu\nu}) =0, \ \ \ \ - \Delta \Phi
+ C R \Phi
+ V'(\Phi )+ \frac{1}{4} f'(\Phi )  F_{\mu\nu}^2 =0, \nonumber \\
&& -\frac{1}{2} (\nabla^{\alpha} \Phi) (\nabla^{\beta} \Phi)
+\frac{1}{4}g^{\alpha\beta} \nabla^{\mu} \Phi \nabla_{\mu}
\Phi +C (\nabla^{\alpha} \nabla^{\beta}- g^{\alpha\beta}
\Delta) \Phi \\
&& +\frac{1}{2} g^{\alpha\beta} V +\frac{1}{8} g^{\alpha\beta} f
F_{\mu\nu}^2
 - \frac{1}{2} f F_{\ \mu}^{\alpha} F^{\beta\mu} =0. \nonumber
\end{eqnarray}
These equations wil be used in the discussion of the on-shell
effective action.

The analysis of one-loop divergences is most conveniently
carried out within the background field method. The fields are
split into their quantum and background parts,
\begin{equation}
\Phi \longrightarrow \bar{\Phi} = \Phi +
\varphi, \ \ \  A_{\mu}  \longrightarrow \bar{A}_{\mu} =A_{\mu} +
Q_{\mu}, \ \ \
g_{\mu\nu} \longrightarrow \bar{g}_{\mu\nu} =g_{\mu\nu}
+h_{\mu\nu},
\end{equation}
where the second terms $ \varphi$, $Q_{\mu}$ and $h_{\mu\nu}$ are
the quantum fields. In what follows we shall use the dynamical
variables $h=g^{\mu\nu}h_{\mu\nu}$ and $\bar{h}_{\mu\nu}
=h_{\mu\nu}-\frac{1}{2}h g_{\mu\nu}$, rather than  $h_{\mu\nu}$.

In order to make contact with Ref. [15] ---where the one-loop
divergences (including the  surface contributions)
in the absence of the Maxwell terms have been calculated--- we add
to the action the following term
\begin{equation}
\Delta S= -\, C \, \xi \int d^2x \, \sqrt{g}  \left[
\Phi \bar{h}_{\mu\nu} (R^{\mu\nu}-
\frac{1}{2} R g^{\mu\nu})
\right],
\end{equation}
where $\xi$ is an arbitrary parameter. Owing to the 2d identity
$R_{\mu\nu}- \frac{1}{2} R g_{\mu\nu}=0$,  expression (4) is
obviously zero. The authors of ref.[15] do not have
a clear interpretation of (4), which
looks very similar to a kind of Wess-Zumino topological term.
(However,they failed to find the classical counterpart for (4).)
Notice that at any stage of the calculation, $\xi$ may be taken to
be equal to zero. Notice also that the second
variation of (4) may be important in order to represent  the
differential operator corresponding to the second variation of the
classical action in a minimal form.Anyway,as we will see $\xi$ (
or more exactly $\gamma$
whatever its origin is) does not appear in
renormalized effective action.

Let us now proceed with the construction of the De Witt covariant
gauge, $\chi^A$, where $A=(*,\mu)$. This gauge will fix the U(1)
transformations and the general covariant transformations
\begin{eqnarray}
\delta_*  Q_\mu =-\nabla_\mu \omega^* \ , \qquad
\delta_* \varphi=\delta_* h=\delta _* \bar h_{\mu \nu }=0, \nonumber \\
\delta Q_\mu =-(\nabla_\nu A_\mu )\omega ^\nu -A_\nu \nabla_\mu
\omega^\nu ,\qquad \delta h=-2\nabla_\mu \omega^\mu, \\
\delta \bar h_{\mu \nu }=-(g_{\mu \lambda} \nabla_\nu +g_{\nu
\lambda }
\nabla_\mu -g_{\mu \nu }\nabla_\lambda )\omega^\lambda,
\qquad \delta \varphi =-(\nabla_\mu \Phi )\omega^\lambda, \nonumber
%\label{}
\end{eqnarray}
where  $\omega_A =(\omega^*,\omega^\mu)$ are parameters of the
gauge transformations.

In order to construct the De Witt covariant gauge we need the metric
on the space of fields $\varphi^i = (Q_\mu,
\varphi,h,\bar{h}_{\mu\nu})$ [24]. It is known that this metric
contains an ambiguity in its definition. In particular, sometimes
this metric is chosen to coincide with a matrix of higher
derivatives of the quadratic expansion of the classical action
[19]. (Notice that this ambiguity leads to a field space metric
dependence of the unique effective action [23].) \, In the case
under discussion the natural choice of this metric $G_{ij}$ is
\begin{equation}
G_{ij}=\sqrt {g}\pmatrix{\Omega (\Phi )g^{\mu \nu } &0 &0 &0 \cr
                  0 & \Theta(\Phi ) &C/2 &0 \cr
                  0 & C/2 &0 &0 \cr
                  0 & 0 & 0 & \gamma \Phi
P^{\mu\nu,\rho\sigma}\cr},
\label{metric}
\end{equation}
where  $\gamma={C\over 2}(\xi -1)$, $ P^{\mu \nu ,\rho \sigma
}=\delta^{\mu \nu ,\rho \sigma }-{1\over
                           2}g^{\mu \nu }g^{\rho \sigma }$,
and the functions $\Omega (\Phi)$ and $ \Theta(\Phi )$ are
arbitrary but sign preserving
(this is the ambiguity in the definition of the
configuration space metric $G_{ij}$).

The De Witt gauge is defined by the following condition
\begin{equation}
{\delta \chi ^A\over  \delta \varphi ^i}=-(c^{-1})^{AB}\nabla^J_B
G_{ij},
\label{DWgauge}
\end{equation}
where the gauge group operators are
\begin{eqnarray}
&\nabla ^{Q_\mu}_*=\nabla _\mu \ ,
\quad \nabla ^{Q_\mu }_\mu =F_{\mu \lambda }+A_\lambda \nabla _\mu,
&\cr
&\nabla ^\varphi _\mu =-(\nabla_\mu \Phi) ,
\quad\nabla^h_\mu =2\nabla _\mu ,
\quad \nabla ^{\bar h_{\mu \nu } }_\lambda=2g_{\lambda \mu }\nabla
_\nu-
g_{\mu \nu }\nabla _\lambda.&
%\label{}
\end{eqnarray}
Notice that the symmmetry group is obviously not the direct product
of the U(1) group and the group of general covariant
transformations, although U(1) is certainly an invariant subgroup
of it. Then, it is natural to choose the non-diagonal matrix
\begin{equation}
c_{AB}={\Omega^2\over f}\sqrt
g \pmatrix{  1& A_\nu\cr  A_\mu&\> A_\mu A_\nu+{2\gamma \Phi f\over
\Omega^2}g_{\mu\nu}\cr},
%\label{}
\end{equation}
which has a Kaluza-Klein structure [24]. (In other words, our
choice of $c_{AB}$ leads to a natural embedding of both
$g_{\mu\nu}$ and $A_\nu$ into a multidimensional metric.) \, Due to
the exact dependence of $c_{AB}$ on $A_\mu$ this matrix is not
covariant. However
\begin{equation}
\det c_{AB}={2\gamma\Phi\Omega^2\over f\, \sqrt{g}}.
%\label{}
\end{equation}
The inverse matrix is given by
\begin{equation}
\left(c^{-1}\right)^{AB}={1\over2\gamma
              \Phi\sqrt g}\pmatrix{
{2\gamma \Phi f\over \Omega^2}+A_{\lambda}A^{\lambda }&\>
-A^{\nu}\cr
-A^{\mu}& g^{\mu\nu}\cr}.
%\label{}
\end{equation}
Finally, from (7), (8), (11), we find
\begin{equation}
S_{GF}=-{1\over 2}\int d^2x\,c_{\hskip
-1ex\phantom{p}_{\scriptstyle AB}}\,
           \chi^A\chi^B,
%\label{}
\end{equation}
where
\begin{eqnarray}
&&\chi ^*=-{f\over \Omega }\nabla ^\mu Q_\mu - {f\over\Omega
^2}\Omega'
             (\nabla ^\mu \Phi )Q_\mu +{\Omega \over 2\gamma
\Phi}A_\lambda
             F^{\mu\lambda } Q_\mu-{\Theta\over2\gamma
\Phi}A^\lambda(\nabla_
             \lambda\Phi )\varphi     \cr\cr
&&\phantom{\chi^*=}+{C\over2\gamma \Phi }A^\lambda \nabla _\lambda
             \varphi-{C\over4\gamma\Phi}A^\lambda(\nabla_\lambda
             \Phi)h+A^\mu\nabla^\nu\bar h_{\mu \nu
}+{1\over\Phi}
             A^\mu(\nabla^\nu\Phi)\bar h_{\mu\nu} \ ,    \\
&&\chi ^\mu ={\Omega \over 2\gamma \Phi }F^{\mu \lambda}Q_\lambda
-{C\over
           2\gamma \Phi }\nabla ^\mu \varphi +{\Theta \over 2\gamma
\Phi }
(\nabla^\mu\Phi)\varphi+{C\over4\gamma\Phi}(\nabla^\mu\Phi)h
\cr\cr
&&\phantom{\chi^\mu=}-\nabla^{\nu}\bar{h}_{\mu\nu}-{1\over\Phi}
(\nabla^{\nu}
           \Phi )\bar{h}^\mu_\nu. \nonumber
%\label{}
\end{eqnarray}
Now the De Witt gauge appears to be minimal. After some tedious
algebra, the total quadratic
expansion of the classical action can be calculated
\begin{eqnarray}
&&S_{tot}^{(2)}= S^{(2)}+ \Delta S^{(2)} + S_{GF} \cr\cr
&&\qquad = -{1\over 2}\int d^2x\, \sqrt g \,\Bigg\{ Q_{\mu}
\Big[-fg^{\mu\alpha}\Delta+\left({f\Omega'\over\Omega}-f'
             \right)(\nabla^{\mu}\Phi)\nabla^{\alpha}
-\left({f\Omega'\over
             \Omega}-f'\right)(\nabla^{\alpha}\Phi)\nabla^{\mu}
 \cr\cr
&&\qquad +\left(2{f{\Omega'}^2\over\Omega
^2}-{f'\Omega'\over\Omega}-{f
\Omega''\over\Omega}\right)(\nabla^{\mu}\Phi)(\nabla^{\alpha}
             \Phi)+{1\over 2}f'g^{\mu\alpha}(\Delta \Phi )
 \cr\cr
&&\qquad +{1\over 2}fRg^{\mu \alpha}+{1\over 2}f''g^{\mu
\alpha}(\nabla^\lambda\Phi)(\nabla_\lambda\Phi)-
{f\Omega'\over\Omega}(\nabla^
             \mu\nabla^\alpha\Phi)+ {\Omega^2\over
2\gamma\Phi}F_{\lambda}^
             {\phantom{\mu}\mu}F^{\lambda\alpha} \Big] Q_\alpha
\cr\cr
&&\ +Q_{\mu}\Big[ \left({\Omega C\over
2\gamma\Phi}+f'\right)F^{\mu\lambda}
             \nabla_\lambda +\left({f'\over2}-{\Omega
C\over4\gamma\Phi}\right)
             (\nabla_\lambda F^{\mu\lambda})      \cr\cr
&&\qquad +\left({f''\over2}+{\Omega
C\over4\gamma\Phi^2}-{C\Omega'\over4\gamma
\Phi}-{\Omega\Theta\over2\gamma\Phi}\right)
F^{\mu\lambda}(\nabla_\lambda\Phi)\Big]\varphi    \cr\cr
&& +\varphi\Big[-\left({\Omega
C\over2\gamma\Phi}+f'\right)F^{\alpha
             \lambda}\nabla_\lambda+\left({f'\over 2}-{\Omega
C\over 4\gamma
             \Phi}\right)(\nabla_\lambda F^{\alpha\lambda})
\cr\cr
&&\qquad +\left({f''\over2}+{\Omega
C\over4\gamma\Phi^2}-{C\Omega'\over4\gamma
             \Phi}-{\Omega\Theta\over2\gamma\Phi}\right)F^{\alpha
\lambda }
             (\nabla_\lambda\Phi)\Big]Q_\alpha        \cr\cr
&&\
+Q_\mu\Big[-{1\over2}fF^{\mu\lambda}\nabla_\lambda-
{1\over4}f(\nabla_\lambda F^{\mu\lambda})-\left({1\over4}f'+{\Omega
C\over4\gamma\Phi}
            \right)F^{\mu\lambda}(\nabla_\lambda\Phi)\Big]h
\cr\cr
&&\ +h\Big[{1\over
2}fF^{\alpha\lambda}\nabla_\lambda-
{1\over4}f(\nabla_\lambda F^{\alpha \lambda})-
 \left({1\over 4}f'+{\Omega
C\over
            4\gamma \Phi}\right)F^{\alpha \lambda} (\nabla_\lambda
\Phi)\Big]
            Q_\alpha    \cr\cr
&&\ +Q_\mu\Big[(\Omega -f)F^{\mu \alpha }\nabla^\beta- fg^{\mu
\beta}
            F^{\alpha \lambda }\nabla_\lambda-{1\over
2}(f+\Omega)(\nabla^\beta F^{\mu \alpha})        \cr\cr
&&\qquad +\left({\Omega\over\Phi }
-{\Omega'\over2}-{1\over2}f'\right)
            F^{\mu\alpha}(\nabla^\beta\Phi)           \cr\cr
&&\qquad -{1\over 2}fg^{\mu\beta}(\nabla_\lambda F^{\alpha
\lambda})-
            {1\over 2}f'g^{\mu
\beta}F^{\alpha\lambda}(\nabla_\lambda \Phi)
            \Big]\bar{h}_{\alpha\beta}             \cr\cr
&&\ +\bar{h}_{\mu \nu }\Big[-(\Omega -f)F^{\alpha\mu}\nabla^\nu
+fg^{\nu
            \alpha}F^{\mu \lambda
}\nabla_\lambda-{1\over2}(f+\Omega)(\nabla^\nu F^{\alpha\mu})
      \cr\cr
&&\qquad +\left(+{\Omega \over \Phi }-{\Omega'\over
2}-{1\over2}f'\right)
            F^{\alpha\mu}(\nabla^\nu\Phi)              \cr\cr
&&\qquad -{1\over 2}fg^{\nu \alpha }(\nabla_\lambda
F^{\mu\lambda})-
            {1\over 2}f'g^{\nu\alpha}F^{\mu \lambda
}(\nabla_\lambda
            \Phi)\Big] Q_{\alpha}                      \cr\cr
&&\
+\varphi\Big[-\left(1+{C^2\over2\gamma\Phi}\right)\Delta+\left({C
\Theta\over
            2\gamma\Phi}-{C^2\over4\gamma\Phi^2}\right)(\Delta\Phi)
   \cr\cr
&&\qquad +\left({C\Theta'\over 2\gamma \Phi}-{C\Theta\over 2\gamma
\Phi^2}+
            {C^2\over 2\gamma \Phi^3}+{\Theta^2\over 2\gamma
\Phi}\right)
            (\nabla_\lambda\Phi)(\nabla^\lambda \Phi)+V''+{1\over
4}f''F^2
            \Big]\varphi                \cr\cr
&&\ +h\Big[{C\over 4}(\Delta \Phi )+{1\over 8}fF^2+{C^2\over8\gamma
\Phi }
            (\nabla^\lambda \Phi)(\nabla_\lambda \Phi)\Big]h
\cr\cr
&&\
+h\Big[-{C\over2}\Delta-{C^2\over4\gamma\Phi}(\nabla^\lambda\Phi)
\nabla_\lambda+\left({C\Theta\over4\gamma\Phi}-{C^2\over8\gamma\Phi^2}
            \right)(\nabla^\lambda\Phi)(\nabla_\lambda\Phi)
\cr\cr
&&\qquad +{C^2\over 8\gamma \Phi} (\Delta \Phi)+{1\over2}V'-{1\over
8}f'
            F^2\Big]\varphi    \cr\cr
&&\
+\varphi\Big[-{C\over2}\Delta+{C^2\over4\gamma\Phi}
(\nabla^\lambda\Phi)
\nabla_\lambda+\left({C\Theta\over4\gamma\Phi}-{C^2\over8\gamma\Phi
            ^2}\right)(\nabla^\lambda\Phi)(\nabla_\lambda\Phi)
\cr\cr
&&\qquad
+{C^2\over8\gamma\Phi}(\Delta\Phi)+{1\over2}V'
-{1\over8}f'F^2\Big]h
\cr\cr
&&\ +\bar{h}_{\mu \nu}\Big[\left(\Theta +{C\over \Phi
}-1\right)(\nabla^\mu
            \Phi)\nabla^\nu +\left({C\over 2\Phi^2}-{\Theta \over
\Phi }+
            {\Theta'\over2}\right)(\nabla^\mu\Phi)(\nabla^\nu\Phi)
  \cr\cr
&&\qquad +\left({\Theta +1\over 2}-{C\over 2\Phi
}\right)(\nabla^\mu
            \nabla^\nu \Phi)-{1\over
2}f'F^{\mu}_{\phantom{\mu}\lambda}
            F^{\nu \lambda}\Big] \varphi              \cr\cr
&&\
+\varphi\Big[-\left(\Theta+{C\over\Phi}-1\right)
(\nabla^\alpha\Phi)\nabla^
\beta+\left({C\over2\Phi^2}-
{\Theta\over\Phi}+{\Theta'\over2}\right)
            (\nabla^\alpha\Phi)(\nabla^\beta\Phi)     \cr\cr
&&\qquad +\left({\Theta +1\over 2}-{C\over 2\Phi
}\right)(\nabla^\alpha
            \nabla^\beta\Phi)- {1\over
2}f'F^{\alpha}_{\phantom{\alpha}\lambda}
            F^{\beta\lambda}\Big]\bar{h}_{\alpha \beta}
\cr\cr
&&\ +\bar{h}_{\mu \nu }\Big[ \left({1\over 4}-{C\over 2\gamma
\Phi}\right)
(\nabla^\mu\Phi)(\nabla^\nu\Phi)+{1\over2}fF^{\mu}_{\phantom{\mu}
            \lambda}F^{\nu \lambda}\Big]h       \cr\cr
&&\ +h\Big[\left({1\over 4}-{C\over 2\gamma
\Phi}\right)(\nabla^\alpha \Phi)
            (\nabla^\beta
\Phi)+{1\over2}fF^{\alpha}_{\phantom{\alpha}\lambda}
            F^{\beta \lambda}\Big]\bar{h}_{\alpha \beta}
\cr\cr
&&\
+\bar{h}_{\mu\nu}\Big[-\gamma\Phi
\delta^{\mu\nu,\alpha\beta}\Delta+\left(
\gamma+{C\over2}\right)g^{\nu\beta}(\nabla^\mu\Phi)\nabla^\alpha
                                                           \cr\cr
&&\qquad -\left(\gamma +{C\over 2}\right)g^{\nu
\beta}(\nabla^\alpha  \Phi)
            \nabla^\mu-\left({3\over 2}C+\gamma\right)g^{\nu
\beta}(\nabla^\mu
            \nabla^\alpha\Phi )                      \cr\cr
&&\qquad +{3\over 4}C(\Delta \Phi ) \delta^{\mu \nu,\alpha \beta}-
            {1\over 4} \delta^{\mu \nu,\alpha \beta}(\nabla_\lambda
\Phi)
            (\nabla^\lambda \Phi)                     \cr\cr
&&\qquad +\left(1+{2\gamma \over \Phi }\right)
g^{\nu \beta}(\nabla^\mu
            \Phi)(\nabla^\alpha \Phi)+\gamma \Phi
R\delta^{\mu\nu,\alpha\beta}
            -{1\over 2}V\delta^{\mu \nu,\alpha \beta}
\cr\cr
&&\qquad -{1\over 8}fF^2\delta^{\mu \nu,\alpha \beta}+g^{\nu \beta
}f
            F^{\mu}_{\phantom{\mu}\lambda}F^{\alpha
\lambda}+{1\over 2}fF^{\mu
            \alpha}F^{\nu \beta}\Big]\bar{h}_{\alpha \beta}
\Bigg\}.
\label{S2tot}
\end{eqnarray}
As is evident from (14), the total quadratic expansion of the
action can be written as follows
\begin{equation}
-\frac{1}{2}\varphi^i\hat{H}\varphi^j\equiv -\frac{1}{2}\varphi^i\left[
\hat{K}_{ij}\Delta+\hat{L}_{\lambda,
ij}\nabla^\lambda+\hat{M}_{ij}\right] \varphi^j,
\end{equation}
where the explicit form of the operator  $\hat{H}$ can readily be
read off from  Eq. (14). However, the extra integrations
by parts change the matrix elements of the operator  $\hat{H}$ and
can destroy its desired properties (properly symmetrized,
$\hat{H}$ should be hermitean). In order to have this operator
uniquely defined, the doubling trick  by 't Hooft and Veltman [18]
is very useful. A clear explanation of how to apply it in the
present context can be found in Refs. [13,14]. In fact, using
this method amounts to the following redefinitions of the operators
in $\hat{H}$ (15):
\begin{eqnarray}
&\hat{H}\to\hat{H}'=-\hat{K}\Delta+\hat{L}'_\lambda\nabla^\lambda
+\hat{M}',  &\cr
&\hat{L}'_{\lambda}={1\over2}\big(\hat{L}_\lambda-
\hat{L}^T_\lambda\big)
                    -\nabla ^\lambda\hat{K} ,
 &\cr
&\hat{M'}={1\over2}\big(\hat{M}+\hat{M}^T\big)-
{1\over2}\nabla^\lambda
          \hat{L}^T_\lambda-{1\over2}\Delta\hat{K},
    &
\label{corrected-matrices}
\end{eqnarray}
where the operators of $\hat{H}'$ are given by
\begin{equation}
\hat{K}_{ij}=\hat{G}_{ij}\Big|_{\phantom{p}_{\Theta\to1+
C^2/2\gamma\Phi}}, \quad
\bigl(\hat K^{-1}\bigr)^{ij}={1\over\sqrt g}\pmatrix
                  {{1\over f}g_{\mu \alpha }&
                  0&0&0\cr 0&0& {2\over C}&0\cr
                  0&{2\over C}&-\left({4\over C^2}+{2\over \gamma
\Phi }
                  \right)&0\cr
                  0&0&0&{1\over \gamma \Phi }P_{\mu \nu ,\alpha
\beta }\cr},
\label{K-1}
\end{equation}
and
\begin{eqnarray}
&&\hat{L'}^\lambda_{11}={f\Omega'-f'\Omega\over\Omega}\left[
                                        (\nabla ^\mu \Phi
)g^{\alpha \lambda }
         -(\nabla ^\alpha \Phi )g^{\mu
\lambda}\right]-f'(\nabla^\lambda\Phi)
         g^{\mu \alpha } \ ;                 \cr\cr
&&\hat{L'}^\lambda_{12}=-\hat{L'}^\lambda_{21}
                                       =\left(f'+{\Omega C\over
2\gamma \Phi }
                                        \right)F^{\mu \lambda}\ ;
   \cr\cr
&&\hat{L'}^\lambda_{13}=-\hat{L'}^\lambda_{31}
                                       =-{1\over 2}fF^{\mu\lambda}\
;   \cr\cr
&&\hat{L'}^\lambda_{14}=-\hat{L'}^\lambda_{41}
=fF^\lambda_{\phantom{\lambda}\omega}
         P^{\alpha \beta ,\mu \omega }+(\Omega
-f)F^\mu_{\phantom{\mu }\omega}
         P^{\alpha \beta ,\lambda\omega } \ ;        \cr\cr
&&\hat{L'}^\lambda_{22}=\left({C^2\over 2\gamma
\Phi^2 }-{C
                                        \Theta \over \gamma \Phi
}\right)
                                        (\nabla^\lambda\Phi) \ ;
  \cr\cr
&&\hat{L'}^\lambda_{23}=-\hat{L'}^\lambda_{32}=
                                        {C^2\over4\gamma \Phi
}(\nabla^\lambda
                                        \Phi ) \ ;  \cr\cr
&&\hat{L'}^\lambda_{24}=-\hat{L'}^\lambda_{42}=
         \left(1-\Theta -{C\over \Phi } \right)(\nabla _\omega \Phi
)P^{\alpha
         \beta ,\lambda \omega }\ ;        \cr\cr
&&\hat{L'}^\lambda_{33}=-{C\over2}(\nabla^\lambda\Phi) \ ;
                                         \cr\cr
&&\hat{L'}^\lambda_{34}=-\hat{L'}^\lambda_{43}=0
                                         \ ;           \cr\cr
&&\hat{L'}^\lambda_{44}=\left({C\over
2}+\gamma\right)(\nabla
                                        ^\omega \Phi )\left[P^{\mu
\nu }_
         {\omega \kappa}P^{\alpha \beta ,\lambda \kappa }-P^{\mu
\nu ,\lambda
         \kappa}P^{\alpha \beta }_{\omega \kappa}\right]-{3C\over
2}(\nabla^
         \lambda\Phi )P^{\mu\nu,\alpha\beta} \ ;      \cr\cr
&&\hat{M'}_{11}={1\over2}fRg^{\mu\alpha}+\left(2{f{\Omega'}^
                                2\over \Omega ^2}-{f'\Omega '\over
\Omega }
         -{f\Omega ''\over \Omega}\right)(\nabla^\mu
\Phi)(\nabla^\alpha\Phi)
                          \cr\cr
&&\phantom{\hat{M'}_{11}=}
-{f\Omega'\over\Omega}(\nabla^\mu
                                           \nabla^\alpha
\Phi)+{\Omega^2\over2
\gamma\Phi}F_\lambda^{\phantom{\lambda}\mu}F_{\lambda\alpha}
         \ ;                \cr\cr
&&               \hat{M'}_{23}=\hat{M'}_{32}=\left({C\Theta
                                \over 4\gamma \Phi
}-{C^2\over8\gamma\Phi^2}
                                \right)(\nabla ^\lambda \Phi
)(\nabla _\lambda
         \Phi)+{C^2\over8\gamma\Phi}(\Delta\Phi)+{1\over
2}V'-{1\over8}f'F^2
             \ ;         \cr\cr
&&\hat{M'}_{33}={C^2\over 8\gamma \Phi
}(\nabla^\lambda\Phi)
                                (\nabla_\lambda\Phi)+{1\over 8}fF^2
\ ; \cr\cr
&&\hat{M'}_{44}=\left[\left(1+{2\gamma \over \Phi
}\right)
(\nabla^\lambda\Phi)(\nabla_\omega\Phi)-\left(
         \gamma+{3C\over
2}\right)(\nabla^\lambda\nabla_\omega\Phi)+fF_{\omega
\rho}F^{\lambda\rho}\right]P^{\mu\nu,\omega\kappa}
P^{\alpha\beta}_{\lambda\kappa}      \cr\cr
&&\phantom{\hat{M'}_{44}=}+\left[-{1\over4}(\nabla^\lambda\Phi)
(\nabla_\lambda\Phi)+\gamma\Phi R-
{1\over2}V-{1\over8}fF^2\right]P^{\mu\nu,\alpha\beta}+{1\over2}f
F^{\omega\lambda}F^{\kappa\rho}P^{\mu\nu}_{\omega\kappa}
P^{\alpha\beta}_{\lambda\rho}.
\end{eqnarray}
The other components of $\hat{M}'$ are not essential for us since
they do not
contribute to the divergencies of the efective action $\Gamma$.

Introducing the notations $\hat{E}^\lambda=-(1/2)\hat{K}^{-
1}\hat{L}^{'\lambda}$ and  $\hat{\Pi}=-\hat{K}^{-1}\hat{M}'$, the
operator $\hat{H}'$ can be put in the form
\begin{equation}
\hat{H}'=-\hat{K} (\hat{1} \Delta+2 \hat{E}^{\lambda}
\nabla_{\lambda} +\hat{\Pi} ).
\end{equation}
The one-loop gravitational-Maxwell contribution to the effective
action is given by the standard expression:
\begin{eqnarray}
&&\Gamma _{div}= \frac{i}{2} \mbox{Tr}\,  \ln \left. \hat{H'}\right|_{div}
= \frac{i}{2} \mbox{Tr}\,  \ln \left.
 (\hat{1} \Delta+2 \hat{E}^{\lambda}
\nabla_{\lambda} +\hat{\Pi} )
\right|_{div} \cr\cr
&& \qquad =
{1\over 2\epsilon}\int\,d^2x\,\mbox{Tr}\, \left[
\hat{\Pi}+{R\over6}\hat1-
\hat{E}^\lambda\hat{E}_\lambda-\nabla_\lambda\hat{E}^\lambda \right]
\label{div-part}
\end{eqnarray}
where $\epsilon =2\pi (n-2)$ and $\mbox{Tr}\,  \ln (-\hat{K})$ gives a
contribution proportional to $\delta (0)$, which is zero in
dimensional regularization. Notice that the term $-
\nabla_{\lambda} \hat{E}^{\lambda} $ is missing in the algorithm
(20) corresponding to Ref. [15]. This will lead to some
disagreement in the surface terms corresponding to the pure
dilatonic sector, as compared with Ref. [15].

The components of  $\hat{E}^\lambda$ and
$\hat{\Pi}$ can be easily evaluated from (18):
\begin{eqnarray}
&&\big(\hat{E}^\lambda\big)^{1}_{1}={f\Omega'-f'\Omega\over
2f\Omega }\left[
(\nabla^\alpha\Phi)g^\lambda_\rho-(\nabla_\rho\Phi)
g^{\alpha\lambda}\right]+{f'\over2f}(\nabla^\lambda\Phi)
         g^\alpha_\rho \ ;             \cr\cr
&&\big(\hat{E}^\lambda\big)^{1}_{2}=-\left({f'\over2f}+{\Omega
C\over 4\gamma
                                    \Phi
f}\right)F_\rho^{\phantom{\rho}\lambda}
                                            \ ;         \cr\cr
&&\big(\hat{E}^\lambda\big)^{1}_{3}={1\over4}
F_\rho^{\phantom{\rho}\lambda};
   \cr\cr
&&\big(\hat{E}^\lambda\big)^{1}_{4}={f-\Omega \over
2f}F_{\rho\omega}
                                    P^{\alpha\beta,\rho \omega
}-{1\over2}
         F^{\lambda\omega}P^{\alpha\beta}_{\rho\omega} \ ;
\cr\cr
&&\big(\hat{E}^\lambda\big)^{2}_{1}=-{f\over2C}F^{\alpha\lambda}\
;    \cr\cr
&&\big(\hat{E}^\lambda\big)^{2}_{2}={C\over 4\gamma \Phi
}(\nabla^\lambda\Phi)
                                         \ ;                 \cr\cr
&&\big(\hat{E}^\lambda\big)^{2}_{3}=\big(\hat{E}^\lambda\big)^{2}
_{4}=0
                                       \ ;          \cr\cr
&&\big(\hat{E}^\lambda\big)^{3}_{1}=\left({f\over
C^2}+{f+\Omega\over2\gamma
                                    \Phi}+{f'\over
C}\right)F^{\alpha\lambda}
                                           \ ;            \cr\cr
&&\big(\hat{E}^\lambda\big)^{3}_{2}=-\left({1\over 2\gamma \Phi
}+{C\over2
                                     \gamma\Phi^2}+{C^2\over
4\gamma^2\Phi^2}
                                     \right)(\nabla^\lambda\Phi) \
;    \cr\cr
&&\big(\hat{E}^\lambda\big)^{3}_{3}=-{C\over 4\gamma \Phi
}(\nabla^\lambda\Phi)
                                           \ ;       \cr\cr
&&\big(\hat{E}^\lambda\big)^{3}_{4}=\left({\Theta-1\over
C}+{1\over\Phi}\right)
(\nabla_\omega\Phi)P^{\alpha\beta,\lambda
                                    \omega}  \ ;       \cr\cr
&&\big(\hat{E}^\lambda\big)^{4}_{1}={f\over 2\gamma \Phi
}F^\lambda_{\phantom
                                    {\lambda }\omega }P^{\alpha
\omega}_{\rho
         \sigma }+{\Omega-f\over 2\gamma \Phi
}F^\alpha_{\phantom{\alpha }
         \omega }P^{\lambda \omega }_{\rho \sigma } \ ;
   \cr\cr
&&\big(\hat{E}^\lambda\big)^{4}_{2}=\left({1-\Theta\over 2\gamma
\Phi }-{C\over
                                    2\gamma \Phi
^2}\right)(\nabla_\omega\Phi)
         P^{\lambda\omega}_{\rho\sigma} \ ;        \cr\cr
&&\big(\hat{E}^\lambda\big)^{4}_{3}=0 \ ;          \cr\cr
&&\big(\hat{E}^\lambda\big)^{4}_{4}=\left({1\over2\Phi}+
{C\over 4\gamma\Phi}
\right)(\nabla^\omega\Phi)\left[P_{\rho
         \sigma}^{\lambda\kappa
}P^{\alpha\beta}_{\omega\kappa}-P_{\rho\sigma,
\omega\kappa}P^{\alpha\beta,\lambda\kappa}\right]+
{1\over 2\gamma\Phi}
  (\nabla^\lambda\Phi)P^{\alpha\beta}_{\rho\sigma} \ ;
\cr\cr
&&\hat{\Pi}^{1}_{1}=-{1\over 2}Rg^\rho _\alpha
+\left({\Omega''\over\Omega}+
                    {f'\Omega'\over
f\Omega}-{2{\Omega'}^2\over\Omega^2}\right)
         (\nabla_\rho\Phi)(\nabla^\alpha\Phi )+{\Omega'\over
\Omega}(\nabla_\rho\nabla ^\alpha\Phi )-{\Omega ^2\over
2\gamma\Phi}F_{\rho\lambda}
         F^{\alpha \lambda } \ ;       \cr\cr
&&\hat{\Pi}^{2}_{2}=-{\Theta\over2\gamma\Phi}(\nabla^\lambda\Phi)
(\nabla_\lambda\Phi)-{1\over C}V'+{f'\over4C}F^2 \ ;
  \cr\cr
&&\hat{\Pi}^{3}_{3}=\left({1-\Theta \over 2\gamma \Phi }+{C\over
2\gamma\Phi^2}
                    +{C^2\over 4\gamma ^2\Phi ^2}
\right)(\nabla^\lambda\Phi)
         (\nabla _\lambda \Phi )+{1\over C}(\Delta\Phi)-{1\over
C}V'  \cr\cr
&&\phantom{\hat{\Pi}^{3}_{3}=}
+\left({f\over2C^2}+{f\over4\gamma\Phi}+{f'
                               \over4C}\right)F^2 \ ;     \cr\cr
&&\hat{\Pi}^{4}_{4}=\left[\left({3C\over 2\gamma \Phi }+{1\over
\Phi } \right)
                    (\nabla^\lambda  \nabla_\omega
\Phi)-\left({1\over \gamma
         \Phi }+{2\over \Phi ^2}\right)(\nabla^\lambda
\Phi)(\nabla_\omega\Phi)
         -{f\over \gamma \Phi
}F_{\omega\nu}F^{\lambda\nu}\right]P^{\omega
         \kappa }_{\rho \sigma }P^{\alpha \beta }_{\lambda\kappa }
    \cr\cr
&&\phantom{\hat{\Pi}^{4}_{4}=}
+\left[\left({1\over2\Phi}-{3C\over4\gamma\Phi}
                               \right)(\Delta \Phi ) +
{1\over4\gamma\Phi}
         (\nabla^\lambda\Phi)(\nabla_\lambda \Phi )-R+{1\over
2\gamma\Phi}V+
         {f\over8\gamma\Phi}F^2\right]P^{\alpha\beta}_{\rho\sigma}
       \cr\cr
&&\phantom{\hat{\Pi}^{4}_{4}=}
-{f\over2\gamma\Phi}F^{\omega\kappa}F^{\lambda
\nu}P_{\rho\sigma,\omega\lambda}P^{\alpha\beta}_{\kappa\nu}.
\end{eqnarray}
Now, it is straightforward to calculate
\begin{eqnarray}
&&\mbox{Tr}\, \big(\nabla_\lambda\hat{E}^\lambda\big)=\nabla_\lambda\left[
\left({f'\over
f}+{1\over\Phi}\right)(\nabla^\lambda \Phi )\right]
    \cr\cr
&&\phantom{\mbox{Tr}\, \big(\nabla_\lambda\hat{E}^\lambda\big)}=\left({f''
\over f}-{{f'}^2 \over
f^2}-{1\over\Phi^2}\right)(\nabla^\lambda\Phi)(\nabla_\lambda
         \Phi)+\left({f'\over f}+{1\over\Phi}\right)(\Delta\Phi)\
;   \cr\cr\cr
&&\mbox{Tr}\, \big(\hat{E}^\lambda\hat{E}_\lambda\big)=\left({f'\Omega'\over
f\Omega}
-{{\Omega'}^2\over2\Omega
^2}-{C\over2\gamma\Phi^2}\right)(\nabla^\lambda\Phi)
(\nabla_\lambda \Phi )       \cr\cr
&&\phantom{\mbox{Tr}\, \big(\hat{E}^\lambda\hat{E}_\lambda\big)=}
+\left({f'\over C}+
{f\over2C^2}+{f\over 4
\gamma\Phi}+{\Omega\over4\gamma\Phi}-{\Omega^2\over2\gamma\Phi f}
         \right) F^2 \ ;           \cr\cr
&&\mbox{Tr}\, \left({R\over6}\hat1+\hat\Pi\right)=-2R-{2\over
C}V'+{1\over\gamma\Phi}V
+\left({f'\over2C}+{f\over2C^2}-
         {\Omega^2\over2\gamma\Phi f}\right)F^2     \cr\cr
&&\phantom{\mbox{Tr}\, \left({R\over6}\hat1+\hat\Pi\right)=}+\left({\Omega
'\over\Omega}
         +{1\over C}+{2\over\Phi}\right)(\Delta\Phi)     \cr\cr
&&\qquad +\left({\Omega''\over\Omega}+{f'\Omega'\over
f\Omega}-{2{\Omega'}^2
\over\Omega^2}-{\Theta\over\gamma\Phi}-{2\over\Phi^2}+{C\over 2
\gamma
\Phi^2}+{C^2\over4\gamma^2\Phi^2}\right)(\nabla^\lambda\Phi)
(\nabla_\lambda\Phi).
\end{eqnarray}
In absence of background vectors ($F^2=0$), Eq. (22) coincides
with the corresponding expression in [15]
(but for not the $\nabla_{\lambda} \hat{E}^{\lambda} $
term, which has been missed in [15]).
Finally, from (20) and (22), we get
\begin{eqnarray}
\Gamma_{GM,div}&=&-{1\over2\epsilon}\int
d^2x\,\sqrt{g}\, \Biggl\{ 2R-{1\over\gamma\Phi}V+{2\over
C}V'+\biggl[{f'
\over2C}+{f\over4\gamma\Phi}+{\Omega\over2\gamma\Phi}\biggr]F^2
\cr\cr
&&{\hskip-20pt} +\biggl[{f'\over f}-{\Omega'\over\Omega}-{1\over
C}-{1\over  \Phi} \biggr](\Delta\Phi)                \cr\cr
&&{\hskip-20pt}
+\biggl[{3{\Omega'}^2\over2\Omega^2}-{\Omega''\over\Omega}-
                {{f'}^2\over f^2}+{f''\over
f}+{\Theta\over\gamma\Phi}+{1\over
\Phi^2}-{C\over\gamma\Phi^2}-{C^2\over4\gamma^2\Phi^2} \biggr]
                (\nabla^\lambda\Phi)(\nabla_\lambda\Phi)\Biggr\}.
\end{eqnarray}
\bigskip

In order to complete this calculation we must take into account the
ghost contribution. The ghost operator $\hat {M}_{gh}$ is defined
as
\begin{equation}
\hat {M}_{gh \ B}^{\ \ A} = \nabla^j_B \, \frac{\delta
\chi^A}{\delta \varphi^j}.
\end{equation}
It has for components
\begin{eqnarray}
&&\hat {M}^*_*={f\over \Omega }\Delta +{f\Omega '\over \Omega
               ^2} (\nabla ^\lambda \Phi )\nabla_\lambda
               +{\Omega\over2\gamma\Phi} A_\mu F^{\mu \lambda }
               \nabla_\lambda,                \cr\cr
&&\hat {M}^\mu
_*=-{\Omega\over2\gamma\Phi}F^{\mu\lambda}\nabla_\lambda\ ,\cr\cr
&&\hat {M}^*_\nu ={f-\Omega \over \Omega }A_\nu
                 \Delta +\left({f\Omega' \over \Omega ^2}-{1\over
\Phi }
                 \right)A_\nu (\nabla^\lambda\Phi) \nabla
_\lambda+{\Omega
                 \over 2\gamma \Phi }A_\mu A_\nu F^{\mu \lambda }
                 \nabla_\lambda                      \cr\cr
&&\phantom{\hat {M}^*_\nu =}\quad
                +{f\over \Omega}(\nabla _\nu A_\lambda )\nabla
^\lambda +
                {f\over \Omega }(\nabla_\lambda A_\nu )\nabla
^\lambda +
                \left({C\over 2\gamma \Phi }+{1\over \Phi }\right)
                A_\lambda (\nabla ^\lambda \Phi )\nabla_\nu
\cr\cr
&&\phantom{\hat {M}^*_\nu =}\quad
                -\left({C\over 2\gamma \Phi }+{1\over \Phi }\right
)A^\lambda
                (\nabla _\nu \Phi )\nabla_\lambda +{f-\Omega \over
2\Omega }
                RA_\nu+{f\over\Omega}(\nabla_\nu\nabla^\lambda
A_\lambda)
\cr\cr
&&\phantom{\hat {M}^*_\nu =}\quad
                +{\Theta \over 2\gamma \Phi }A_\lambda (\nabla
^\lambda\Phi)
                (\nabla _\nu \Phi )-{C\over 2\gamma \Phi }A_\lambda
                (\nabla ^\lambda \nabla _\nu \Phi )+{f\Omega'\over
\Omega ^2}
                (\nabla ^\lambda \Phi )(\nabla _\nu A_\lambda )
\cr\cr
&&\phantom{\hat {M}^*_\nu =}\quad
                +{\Omega \over 2\gamma \Phi }A_\mu F^{\mu
                \lambda } (\nabla _\nu A_\lambda ),
\cr\cr
&&\hat{M}^\mu_\nu=\delta ^\mu _\nu \Delta +{1\over \Phi }\delta
^\mu _\nu
                 (\nabla _\lambda \Phi )\nabla ^\lambda -{\Omega
\over 2\gamma
                 \Phi }A_\nu F^{\mu \lambda }\nabla ^\lambda +
                 \left({C\over 2\gamma \Phi }+{1\over \Phi }
\right)(\nabla_\nu\Phi)\nabla ^\mu                    \cr\cr
&&\phantom{\hat{M}^\mu_\nu=}\quad
                - \left({C\over 2\gamma \Phi }+{1\over \Phi }
\right)(\nabla^\mu\Phi)\nabla _\nu+{R\over 2}\delta^\mu _\nu
+{C\over 2 \gamma \Phi }(\nabla ^\mu \nabla _\nu \Phi )
  \cr\cr
&&\phantom{\hat{M}^\mu_\nu=}\quad
                  -{\Theta \over 2\gamma \Phi }(\nabla^\mu\Phi
)(\nabla _\nu
                 \Phi)-{\Omega \over 2\gamma \Phi }F^{\mu \lambda
}(\nabla _\nu
                 A_\lambda ).
%\label{}
\end{eqnarray}
As mentioned, the ghost operator is not manifestly U(1)
gauge-covariant
though $\ln\det\hat M$ certainly is.

Now we follow the usual way of calculating the $\ln\det$. (Clearly,
no doubling trick is needed, because ghost operator is simply given
by (24).) Define:
\begin{equation}
\hat{M}^A_B=\hat{K}^A_B \Delta
+\hat{L}^\lambda\nabla_\lambda+\hat{P}^A_B \end{equation}
so that
$$
\bigl(\hat{K}^{-1}\bigr)^A_B=\pmatrix{{\Omega\over
f}&{\Omega-f\over f}A_\nu\cr
                                   0&\delta ^\mu _\nu \cr},
 \qquad\det\hat K ={f\over \Omega},
$$
$$
\mbox{Tr}\, \hat{1}=1+\delta ^\mu _\mu =3,
$$
and the explicit form of $\hat{L}^\lambda$ and $\hat{P}$ follows from
(25).
We introduce the matrices
\begin{equation}
\hat E={1\over 2}\hat{K}^{-1}\hat{L}^\lambda , \hat
\Pi=\hat{K}^{-1}\hat P
\end{equation}
and make use of Eq. (20) to get
\begin{eqnarray}
&&\Gamma _{gh,div}= -i \mbox{Tr}\,  \ln \hat{M}_{gh} =
 -{1\over 2 \epsilon }\int d^2 x\,\sqrt
g \, \Biggl\{
                    3R-{\Omega \over 2\gamma \Phi
}F^2+\biggl[{C\over \gamma
                   \Phi }-{2\over\Phi}
-{\Omega'\over\Omega}\biggr](\Delta\Phi)
\cr\cr
&&\phantom{\Gamma _{gh,div}=}+\biggl[{{\Omega'}^2\over 2\Omega
^2}-{\Omega''
                   \over\Omega}-{\Theta \over \gamma \Phi }+{2\over
\Phi ^2}
                   + {C^2\over 4\gamma ^2\Phi ^2}+{C\over \gamma
\Phi ^2}
\biggr](\nabla_\lambda\Phi)(\nabla^\lambda\Phi)\Biggr\}.
%\label{}
\end{eqnarray}
The total divergent part of the effective action in the De Witt
gauge is given by the sum of (23) and (28):
\begin{eqnarray}
\Gamma_{div}=&& -{1\over 2\epsilon}\int d^2x\,\sqrt{g}\,\Biggl\{
5R- {1\over\gamma\Phi}V+{2\over
C}V'+\biggl[{f'\over2C}+{f\over4
                \gamma\Phi}\biggr]F^2         \cr\cr
&& +\biggl[{f'\over
f}-2{\Omega'\over\Omega}+{C\over\gamma\Phi}-{1\over C}-{3
   \over\Phi}\biggr](\Delta\Phi)      \cr\cr
&&
+\biggl[2{{\Omega'}^2\over\Omega^2}-2{\Omega''\over\Omega}
-{{f'}^2\over f^2}
   +{f''\over f}+{3\over\Phi^2}
\biggr](\nabla^\lambda\Phi)(\nabla_\lambda\Phi)
   \Biggr\}.
\label{gammadiv}
\end{eqnarray}
This expression constitutes the main result of the present section.

A few remarks are in order. First of all, after dropping the
surface divergent terms (which are kept in (29)) and putting
$F^2=0$, the result (29) agrees with the calculations done in Ref.
[15], and for $\xi =0$ ($\gamma = -C/2$) with the result of Refs.
[13,14] in the same gauge. The divergences of the Maxwell sector
coincide with the results of Ref. [16], and for $\xi =0$ with those
of Ref. [11]. As we can see, the Maxwell sector is
$\Omega$-independent. (In fact, the Maxwell sector looks the same
in the three different covariant gauges, as it will be discussed
below.)

Moreover, $\Gamma_{div}$ does not depend on $\Theta$, in accordance
with the general results of Ref. [14] (notice that $\Theta$ is a
particular case of the function $X(\Phi )$ introduced in Ref.
[14]).

Before discussing renormalization, let us perform the calculation
of  $\Gamma_{div}$ in a different covariant gauge, which we here
call $\Omega$-degenerate De Witt gauge. It is chosen as
\begin{equation}
\chi ^*=-\nabla^\mu Q_\mu
%\label{}
\end{equation}
in the Maxwell sector  and the covariant De Witt gauge in the pure
gravitational sector [15]:
\begin{equation}
\chi ^\mu
=-{C\over2\gamma\Phi}\nabla^\mu\varphi+{\Theta\over2\gamma\Phi}
(\nabla^\mu\Phi)\varphi+{C\over4\gamma\Phi}(\nabla^\mu\Phi)h-
\nabla^\nu\bar{h}_{\mu\nu}-{1\over\Phi}(\nabla^\nu\Phi)
\bar{h}^\mu_\nu.
%\label{}
\end{equation}
The matrix $c_{AB}$ is chosen to be
\begin{equation}
c_{AB}=\sqrt{g}
\pmatrix{f&0 \cr0&2\gamma\Phi g_{\mu\nu} \cr}.
%\label{}
\end{equation}
The calculation can be done in direct analogy with the above case.
The gravitational-Maxwell contribution to $\Gamma_{div}$ is given
by (23) with all $\Omega$-terms dropped. The ghost operator is
again non-diagonal:
\begin{eqnarray}
&&\hat M^*_*=\Delta \ , \qquad\qquad  \hat M^\mu _*=0 \ ,   \cr\cr
&&\hat M^*_\nu =A_\nu\Delta +(\nabla_\lambda A_\nu)\nabla^\lambda
+  (\nabla_\nu A_\lambda )\nabla^\lambda
+(\nabla^\lambda
              \nabla_\nu A_\lambda) \ ,        \cr\cr
&&\hat M^\mu _\nu =\delta^\mu_\nu\Delta +{1\over\Phi
}(\nabla^\lambda \Phi )
                \delta^\mu _\nu \nabla_\lambda
+\bigl({C\over2\gamma \Phi }
               +{1\over\Phi }\bigr)\nabla^\mu     \cr\cr
&&\phantom{M^\mu _\nu=}\quad -\bigl({C\over2\gamma \Phi
}+{1\over\Phi }\bigr)+
                {R\over2}\delta^\mu _\nu +{C\over2\gamma \Phi
}(\nabla^\mu
                \nabla_\nu \Phi )-{\Theta \over2\gamma \Phi
}(\nabla^\mu \Phi )
                (\nabla_\nu \Phi ).
%\label{}
\end{eqnarray}
Using (33) we can find the ghost contribution in the form
\begin{eqnarray}
&&\Gamma_{gh,div}=-{1\over 2\epsilon}\int d^2 x\,\sqrt
g\,\Biggl\{3R+\biggl[
{C\over\gamma\Phi}-{2\over\Phi}\biggr](\Delta\Phi)   \cr\cr
&&\phantom{\Gamma_{gh,div}=}\qquad
+\biggl[{C^2\over4\gamma^2\Phi^2}+{C\over
\gamma\Phi^2}+{2\over\Phi^2}-{\Theta\over\gamma\Phi}\biggr]
                   (\nabla^\lambda\Phi)(\nabla_\lambda\Phi)
\Biggr\}.
%\label{}
\end{eqnarray}
Summing up, we find that the divergent part of the effective action
in the $\Omega$-degenerate De Witt gauge is given by Eq. (29)
discarding all $\Omega$-dependent terms.(It does not mean that
$\Omega$ =0.Rather,one sets $\Omega$= const and after that
setting it equal to zero .Otherwise,configuration space
metric diverges.)
 This justifies the
nickname ``$\Omega$-degenerate'' De Witt gauge given to the gauge
(30)-(31).

For completeness, we shall now write $\Gamma_{div}$ in the simplest
covariant gauge [11,16]:
\begin{equation}
\chi ^*=-\nabla^\mu Q_\mu, \ \ \
\chi ^\mu
=-{C\over2\gamma\Phi}\nabla^\mu\varphi-
\nabla^\nu\bar{h}_{\mu\nu},
%\label{}
\end{equation}
with the same $c_{AB}$ (32). An explicit evaluation in this case
yields [16]:
\begin{eqnarray}
&&\Gamma_{div}=-{1\over 2\epsilon}\int d^2 x\,\sqrt g\,
\Biggl\{5R-{1\over
                \gamma\Phi}V+{2\over
C}V'+\biggl[{f\over4\gamma\Phi}+{f'\over
                2C} \biggr] F_{\mu\nu}^2
\cr\cr &&\phantom{\Gamma_{div}=}\quad+\biggl[{f'\over f}
+ {1\over \Phi}- {1\over
C} \biggr] \Delta\Phi +\biggl[{f''\over
f}-{{f'}^2 \over
f^2}-{1\over\Phi^2}+ {C\over \gamma \Phi^2}\biggr](\nabla^\lambda\Phi)
 (\nabla_\lambda\Phi)\Biggr\}.
%\label{}
\end{eqnarray}
Thus we have calculated the one-loop divergences of the convenient
effective action in three different covariant gauges: (i) the De
Witt gauge, (ii) the $\Omega$-degenerate De Witt gauge, and (iii)
the simplest covariant gauge. We see that $\Gamma_{div}$ is
explicitly gauge dependent. Moreover, when the term (4) is present,
$\Gamma_{div}$ depends on the parameter $\gamma$.

Let us now discuss the renormalizability of the theory off-shell.
By dropping the surface terms in  $\Gamma_{div}$ we see that all
three gauges under consideration lead to the same off-shell
effective action:
\begin{equation}
\Gamma_{div}=-{1\over 2\epsilon}\int d^2 x\,\sqrt g\,
\Biggl\{{2\over
C}V'-\frac{V}{\gamma \Phi}+\biggl[{f'\over
                2C}+ \frac{f}{4\gamma \Phi} \biggr] F_{\mu\nu}^2
+{C\over\gamma\Phi^2}(\nabla^\lambda\Phi)
 (\nabla_\lambda\Phi)\Biggr\}.
\end{equation}
Adding to the classical action (1) the corresponding counterterms
((37) with the opposite sign) we obtained the renormalized action.
Choosing the one-loop renormalization of $g_{\mu\nu}$ as
\begin{equation}
g_{\mu\nu}= \exp \left( - \frac{1}{2\epsilon\gamma \Phi} \right)
\widetilde{g}_{\mu\nu},
\end{equation}
we get the one-loop renormalized action in the following form
\begin{equation}
S_R = -\int d^2 x\,\sqrt{\widetilde{g}} \,
\left[ \frac{1}{2} \widetilde{g}^{\mu\nu} \partial_\mu \Phi \partial_\nu
\Phi+ C \widetilde{R} \Phi + {V}  -
  \frac{V'}{\epsilon C} +  \frac{1}{4} \left(
1- \frac{f'(\Phi)}{4\epsilon C} \right) \widetilde{g}^{\mu\alpha}
\widetilde{g}^{\nu\beta}  F_{\mu\nu} F_{\alpha\beta} \right].
\end{equation}
The dilaton and the coupling $C$ do not get renormalized in the
one-loop approximation.

It follows from (39) that  the theory under discussion is one-loop
multiplicatively renormalizable for the families of potentials:
\begin{eqnarray}
V(\Phi)= e^{\alpha \Phi} + \Lambda, & & f(\Phi)= e^{\beta \Phi} +
f_1, \nonumber \\
V(\Phi)= A_1 \sin \Phi + B_1 \cos \Phi & & f(\Phi)= A_2 \sin \Phi
+ B_2 \cos \Phi,
\end{eqnarray}
where $\alpha$, $\Lambda$, $\beta$, $f_1$, $A_1$, $B_1$, $A_2$ and
$B_2$ are arbitrary couplings. Let us recall that the black hole
solutions for the Liouville like potentials of (40) have been
studied in [4-7] in the case of dilaton gravity with $N$ scalars,
and in [9,11] in the case of 2d dilaton-Maxwell gravity. Notice
also the fact that, in contradistinction with 4d Einstein-Maxwell
theory ---which is not renormalizable [17]--- in the theory under
discussion we get off-shell renormalizability.

As we can see from the above calculations, in the three different
gauges considered $\Gamma_{div}$ is given by three different
expressions: the convenient effective action is gauge dependent.
However, all the differences are contanined in the surface
counterterms only. If we drop surface terms we surprisingly find
that the effective action is the same
in the three gauges considered.
 Hence, due to some reason ---and at least in the gauges
under discussion here--- the gauge dependent divergences of the
one-loop effective action in 2d dilaton-Maxwell gravity are
included in the surface divergences.

Let us now study the on-shell limit of the effective action.
Starting from (29), integrating by parts, keeping all the surface
terms, and using the second and the third of the classical field
equations (2), we get the on-shell divergences of the effective
action:
\begin{equation}
\Gamma_{div}^{on-shell}=-{1\over 2\epsilon}\int d^2 x\,\sqrt{g} \,
\left\{ 3R + \Delta \left[ \ln \left( \frac{f }{\Phi^3} \right) -
\ln \Omega^2 + \frac{1}{C} \Phi \right] \right\}.
\end{equation}
Hence, one can see that the one-loop divergences of the on-shell
effective action are just given by surface terms. In other words,
the one-loop $S$-matrix is finite as in pure dilaton gravity. For
comparison, remember that in 4d Einstein-Maxwell theory the
one-loop $S$-matrix is not finite [17], while it is so in
 4d Einstein theory [18]. The other interesting point to be noticed
concerns the arbitrary gauge function $\Omega (\Phi)$. The fact
that this function is present in (41) explicitly shows that
on-shell surface divergences are gauge dependent. Indeed, had we
started from $\Gamma_{div}$ in the $\Omega$-degenerate De Witt
gauge, we would have got on-shell the same expression (41) without
$\Omega$ terms. Using  $\Gamma_{div}$ (36) on shell again leads to
an expression different from (41) because of some surface
divergences.

\section{One-loop unique effective action divergences in 2d
dilaton-Maxwell gravity}

In this section we will study the one-loop unique effective action
for 2d dilaton-Maxwell gravity. As is well known, this action is
gauge invariant, gauge fixing and reparametrization independent.
Actually, there is a whole family of unique effective actions
[19-21]. However, all members of the family coincide in the
one-loop approximation to quantum gravity, so there is no need to
discuss here these differences in the definition of the action
(review articles on the unique effective action are listed in Ref.
[21]). Notice, however, that the unique effective action is
configuration space metric dependent [23]. This is why, actually,
the unique effective action does not solve the gauge dependence
problem of the convenient effective action (as it had been claimed
in the first works [19,20]), and the nickname ``unique'' does not have
a proper sense. In fact we have a {\it gauge} dependence of the
convenient effective action versus a {\it configuration space metric}
dependence of the unique effective action. The unique effective
action is still a useful covariant formalism which can add some
information to the convenient effective action formalism. Moreover,
there is still a hope to construct a physical off-shell effective
action which would be really unique, along the direction started in
[19,20].     Hence, the
discussion of the unique effective action in situations where it
can be compared with the convenient effective action is very
useful.
The finite parts of the unique effective action in 2d gravity have
been considered in Refs. [22]. The one-loop divergences in pure
dilaton gravity have been obtained in Ref. [15]. In what follows,
we will generalize the calculations of Ref. [15] to the case of
dilaton-Maxwell gravity.

According to Refs. [19-21], one should add to the total quadratic
expansion of the convenient effective action (19) in De Witt's
gauge the correction
\begin{equation}
-\frac{1}{2} \Gamma_{jk}^i \varphi^j \varphi^k \frac{\delta
S}{\delta  \varphi^i},
\end{equation}
where $ \Gamma_{jk}^i = \left\{_{jk}^{ \ i} \right\} + T_{jk}^i$
[19-21] is the connection on the space of fields. This procedure
will give the one-loop unique effective action. (Notice that as a
consequence of the classical equations of motion the correction
(42) is zero on-shell.)

Now we proceed with the calculation of the corrections introduced
by (42). Notice that it is more convenient to use in what follows
the dynamical variable $h_{\mu\nu}$, and not  $\bar{h}_{\mu\nu}$
and $h$. The index $i$ in (42) runs through $\{ Q_{\mu}, \varphi,
h_{\mu\nu} \}$. For the configuration space metric (6), the
Christoffel symbols $\left\{_{jk}^{ \ i} \right\}$ can be easily
calculated. Part of them are listed in Ref. [15] (Eq. (26)); those
do not change. We will write below the remaining non-zero
components which appear in dilaton-Maxwell gravity:
\begin{eqnarray}
&&\left\{\matrix{Q_\alpha \cr\varphi\ Q_\mu \cr}\right\}=
                   {\Omega '\over 2\Omega }\delta ^\mu _\alpha,  \
\ \ \
\left\{\matrix{Q_\alpha \cr Q_\mu \ h_{\rho \sigma }\cr}\right\}=
               -{1\over
2}g_{\lambda\alpha}P^{\lambda\mu,\rho\sigma}, \cr\cr
&&\left\{\matrix{h_{\mu \nu }\cr Q_\alpha \ Q_\beta \cr}\right\}=
               {\Omega \over 2\gamma \Phi }P^{\alpha \beta }_{\mu
\nu }   -{\Omega '\over 2C}g_{\mu \nu }g^{\alpha \beta } .
\label{Crist}
\end{eqnarray}
Now, as has been discussed in [15] the local correction is given by
\begin{equation}
\Gamma_{loc}^{VD}=-{1\over 2 \epsilon }\int d^2x\,\sqrt
g\,\left[
       \left(\hat{K}^{-1}\right)^{ij}\left\{{k\atop i\
j}\right\}S,_k\right],
%\label{}
\end{equation}
where $\hat {K}^{-1}$
is the same as before (see Eq. (17)
but in new
variables. In our case we easily get
\begin{eqnarray}
&&\Gamma_{loc}^{VD}=-{1\over 2\epsilon}\int d^2 x\,\sqrt
g\,\Biggl\{
                 R+{1\over C}V'+\biggl[ {1\over \gamma \Phi
}+{1\over C\Phi }
                 -{2\Theta\over C^2}+{\Omega'\over Cf}\biggr]V
 \cr\cr
&&\phantom{\Gamma_{loc.VD,div}=-}\quad +\biggl[{2\Theta \over
C}-{1\over C}-
                 {1\over \Phi }-{C\over\gamma\Phi}-{\Omega'\over
f}\biggr](
                 \Delta\Phi)                      \cr\cr
&&\phantom{\Gamma_{loc.VD,div}=-}\quad +\biggl[{\Theta \over
2C^2}-{1\over 4
                 \gamma \Phi }-{1\over 4C\Phi }+{f'-\Omega '\over
4Cf}\biggr]f
                 F^2 \Biggr\}.
\label{gammalocal}
\end{eqnarray}

Now let us proceed with the evaluation of the non-local
 Vilkovisky-De Witt correction connected with the torsion
$T_{jk}^i$ in (42). The generalized Schwinger-De Witt technique
(see [26] for details) is very useful in this case. The application
of this technique is based on having a De Witt covariant gauge,
what we actually do (Sect. 2). As it has been discussed in detail
in Ref. [15], the torsion (non-local) Vilkovisky-De Witt correction
to the effective action in De Witt's gauge is given by
\begin{equation}
\Gamma _{\tau ,div}=-{i\over 2}\mbox{Tr}\, \hat U_1|_{div},
\end{equation}
where
\begin{eqnarray}
\hat {U}^A_{1B}&=&\hat N^{AA'}\nabla^i_{A'}\left({\cal D}_i\nabla
^j_{A''} \right)
             S,_j   \hat{N}^{A'B'}c_{\vphantom{p}_{\scriptstyle
B'B}},
\  \ \  \hat N^{AB}=\left(\hat N_{AB}\right)^{-1}, \nonumber \\
 \hat{N}_{AB}& =& -C_{AA'} \hat{M}^{A'}_{gh \, B} =-\Omega \sqrt{g}
\left( \begin{array}{cc} 1  & A_{\mu} \\ A_{\nu} & A_\mu A_\nu + 2\gamma
\Phi/\Omega \end{array} \right) \Delta + \cdots
\end{eqnarray}
where $\hat{M}_{gh}$ and $c_{AB}$ are given by (25) and (9),
respectively, $\nabla^j_B$ are the gauge generators, ${\cal D}_i$
the covariant derivative in the space of fields (${\cal D}_i$ is
constructed with the Christoffel symbols), and the lower derivative
terms in the third expression (47) may be discarded.

Introduce the notations
\begin{eqnarray}
&&c_{A(x)B(y)}\equiv c_{AB}(x)\,\delta (x-y), \nonumber \\
&&\nabla ^{i(x)}_{A(y)}=t^i_A\nabla _x\delta (x-y)+\dots, \nonumber \\
&&\hat{N}^{A(x)B(y)}={\cal
N}^{AB}(x){1\over\Delta_x}\delta(x-y)+\dots, \\
&&{\cal D}_ {i(x)}\nabla ^{j(y)}_{A(z)}={\cal D}^j_{iA} (y)\delta
(y-z)\nabla_y \delta (y-x)+ \cdots, \nonumber
\end{eqnarray}
where the lower order derivatives may again be omitted. Then,
the divergent structure of $\hat{U}_1$ becomes evident (see Ref.
[15])
\begin{eqnarray}
&&\hat{U}^{A(x)}_{1 \, B(y)}={\cal U}^A_B (x)\nabla_x \nabla_x {1\over
                       \Delta_x^2 } \delta (x-y)+\dots, \nonumber \\
&&{\cal U}^A_B=-{\cal N}^{AA'}t^i_{A'}{\cal D}^j_{i\,A''}S,_j {\cal
N}^{A''B'}
              c_{\vphantom{p}_{\scriptstyle B'B}},
\end{eqnarray}
and [26]
\begin{equation}
\nabla_{\mu (x)} \nabla_{\nu (x)}{1\over\Delta_x^2 }\delta (x-y)
\bigg|_{y\to x}\longrightarrow
-{i\over 2\epsilon}\sqrt{g(x)}\,g_{\mu\nu}(x).
%\label{}
\end{equation}
The functional trace in (46)
\begin{equation}
\mbox{Tr}\, \ldots \equiv \int\!\! d^2x\! \sum_{A} \lim_{y\to x}\dots,
%\label{} {A\atop A=B}
\end{equation}
can be performed to yield
\begin{eqnarray}
&&{\cal U}^A_A=-R^{AB}t^i_B{\cal D}^j_{i\,A}S,_i, \\
&& R^{AB}={\cal N}^{AA'}c_{\vphantom{p}_{\scriptstyle A'B'}} {\cal
N}^{B'B}={1\over 2\gamma \Phi \sqrt g}\pmatrix{
{2\gamma \Phi \over f}+A^\lambda A_\lambda &-A^\nu \cr
-A^\mu& g^{\mu \nu }\cr}. \nonumber
\end{eqnarray}
Calculating the covariant derivatives and using (46) and (49)-(52),
we obtain
\begin{eqnarray}
&&\Gamma _{\tau ,div}=-{1\over 2 \epsilon }\int d^2x\,\sqrt g\,
\Biggl\{
                  {1\over C\sqrt g}{\delta
S\over\delta\Phi}+\left({\Omega'
                  \over 2Cf}+{1\over C\Phi }-{\Theta \over C^2}
\right){1\over
                  \sqrt g}g_{\mu\nu}{\delta S\over\delta
g_{\mu\nu}}\Biggr\} \cr\cr
&&=-{1\over 2\epsilon}\int d^2x\,\sqrt
g\,\Biggl\{-R-
                      {1\over C}V'+\biggl[ {\Theta \over
C^2}-{1\over C\Phi }-
                      {\Omega '\over2Cf} \biggr]V            \cr\cr
&&\phantom{\Gamma _{\tau ,div}=}\quad +\biggl[{\Omega '-2f'\over
8C}+{f\over
                      4C\Phi }-{f\Theta \over 4C^2}
\biggr]F^2+\biggl[{\Omega'
                      \over 2f}+{1\over C}+{1\over\Phi
}-{\Theta\over C}\biggr]
                      (\Delta \Phi )\Biggr\}.
%\label{}
\end{eqnarray}
The total Vilkovisky-De Witt correction to the divergent part of
the effective action (29) is given by the sum of (45) and (53).
Notice that this sum vanishes on shell and that, discarding the
Maxwell sector, it coincides with the result obtained in Ref. [15].

Finally, the one-loop unique effective action is given by the sum
of (29), (45), and  (53):
\begin{eqnarray}
\Gamma _{unique,div}=&& -{1\over 2\epsilon}\int
d^2x\,\sqrt{g}\,\Biggl\{ 5R
                      +{2\over
C}V'+\biggl[{\Omega'\over2Cf}-{\Theta\over C^2}
                      \biggr]V                      \cr\cr
&&+\biggl[{f'\over2C}-{\Omega'\over8C}+{f\Theta\over4C^2}\biggr]F^2
    \cr\cr
&& +\biggl[{f'\over
f}-2{\Omega'\over\Omega}-{\Omega'\over2f}+{\Theta\over C}
-{1\over C}-{3\over\Phi}\biggr](\Delta\Phi)     \cr\cr
&&
+\biggl[2{{\Omega'}^2\over\Omega^2}-2{\Omega''\over\Omega}-{
{f'}^2\over f^2}
   +{f''\over f}+{3\over\Phi^2}
\biggr](\nabla^\lambda\Phi)(\nabla_\lambda\Phi)
   \Biggr\}.
%\label{}
\end{eqnarray}
It is evident that the unique effective action (54) on-shell leads
to the same expression (41) as the convenient effective action
(29). The configuration space metric dependence (through the
arbitrary functions  $\Theta (\Phi)$ and $\Omega(\Phi)$
is seen explicitly in (54).

What about the renormalization of the unique effective action
off-shell? Choosing the renormalization of $g_{\mu\nu}$ as
\begin{equation}
g_{\mu\nu}= \exp \left[ \frac{\Omega'}{4C\epsilon f}-
\frac{\Theta}{2C^2\epsilon} \right] \widetilde{g}_{\mu\nu},
\end{equation}
we get the renormalized effective action in the form
\begin{eqnarray}
S_R &=&- \int d^2 x\,\sqrt{\widetilde{g}} \,
\left[ \frac{1}{2} \widetilde{g}^{\mu\nu} \partial_\mu \Phi \partial_\nu
\Phi+ C \widetilde{R} \Phi +V(\Phi) -  \frac{V'}{\epsilon C}
 \right. \nonumber \\
&+& \left. \frac{1}{4} \left(
1- \frac{f'(\Phi)}{\epsilon C f(\Phi)} \right) \widetilde{g}^{\mu\alpha}
\widetilde{g}^{\nu\beta}  F_{\mu\nu} F_{\alpha\beta} \right].
\end{eqnarray}
Thus, the one-loop renormalization of the metric tensor (55) in
this case is different from the one corresponding to the convenient
effective action formalism. However, the renormalized effective
action (56) looks exactly the same, and leads to the same class of
renormalizable dilaton and Maxwell potentials as given in (40).

\section{Conclusions}

Summing up, we have discussed the one-loop renormalization
structure of 2d dilaton-Maxwell gravity. The one-loop convenient
effective action has been found in three different covariant gauges.
However, the gauge dependence appears only in surface divergent
terms. Moreover, the on-shell effective action is given by surface
divergent terms only (on-shell finiteness); however, these terms
are still gauge dependent. The one-loop divergences of the unique
effective action ---which is known to be gauge fixing independent and
parametrization independent--- have been found. This effective action
is explicitly demonstrated to be configuration space metric
dependent.

The off-shell renormalizability  leads to the same renormalizable
dilaton and Maxwell potentials both for the case of the convenient
and for the  unique effective action. It is very interesting to
notice that the heterotic string effective action is described as
just a particular case of the renormalizable dilaton-Maxwell theory
(1). It would be important to examine 2d quantum cosmology [30] for
this renormalizable model.

The other topic which deserves further study has to do with the
generalization of the model to include other dilaton-matter
systems. As we show in the Appendix for the case of the 2d
dilaton-Yang-Mills theory, there appears only a trivial (surface
divergence) correction to the one-loop effective action, as
compared with the dilaton-Maxwell theory. Hence, there are no loop
corrections from the gravitational coupling $C$ to Yang-Mills
coupling (which is dimensional). However, if  we add fermions
to the theory, we can expect such corrections to appear in the
four-fermion coupling [28] and in the Pauli coupling [29]. Both
these coupling constants are dimensionless and this would not be in
contradistinction with renormalizability. Moreover, the chiral anomaly
structure can be very interesting in this case. Work along this
line is in progress.
\vspace{5mm}

\noindent{\large \bf Acknowledgments}

We would like to thank I. Antoniadis, F. Englert, R. Kantowski, C.
Marzban, F. Mazzitelli, N. Mohammedi, A.A. Slavnov, and I.V. Tyutin
for useful discussions at different stages of this work.
S.D.O. wishes to thank the Japan Society for the Promotion of
Science (JSPS, Japan) for financial support and T. Muta and the
Particle Physics Group at Hiroshima University for kind
hospitality.
E.E. has been supported by DGICYT (Spain), research project
PB90-0022, and by the Alexander von Humboldt Foundation (Germany).
\bigskip

\appendix
\section{Appendix}

In this Appendix we extend the calculations of Sect. 2 to the case
of dilaton-Yang-Mills gravity. We start again from action (1) with
the only change of $F_{\mu\nu}$ by $F^a_{\mu\nu}$, where
\begin{equation}
F^a_{\mu\nu}=\nabla_\mu A_\nu^a-\nabla_\nu
A_\mu^a+ f^a_{\phantom{a}bc}A_\mu^b A_\nu^c,
\end{equation}
being $f_{abc}$ antisymmetric, and the gauge group is chosen to be
simple and compact (for a recent discussion of 2d quantum gauge
theory see [27] and references therein).

The background field method will be employed, exactly as in Sect.
2 (only $Q_\mu$ must be substituted by  $Q^a_\mu$) and the simplest
covariant gauge is considered, where
\begin{equation}
c_{AB}=\sqrt
g\pmatrix{f\gamma_{ab}&0\cr
                                                  0&2\gamma\Phi
g_{\mu\nu}\cr},
\end{equation}
and $\gamma_{ab}=-f_{ad}^c f_{bc}^d$ is the  positive Killing
metric. The calculation can be repeated along the same lines as in
Sect. 2 (see also Ref. [16]). Actually, only the $QQ$-sector is
going to change. In comparison with [16], the following components
of $\hat{E}^\lambda$ and  $\hat{\Pi}$ are different
\begin{eqnarray}
&&\bigl(\hat{E}^\lambda\bigr)^1_1={f'\over2f}\delta^a_b
\left[g^{\alpha\lambda}
(\nabla_\rho\Phi)-g^\lambda_\rho(\nabla^
         \alpha\Phi)+g^\alpha_\rho(\nabla^\lambda\Phi)\right],
    \cr \cr
&&\hat{\Pi}^1_1=2f^a_{\phantom{a}cb}F_\rho^{c\,\alpha}-\delta^a_b
R^\alpha_\rho.
%\label{}
\end{eqnarray}
Finally, we obtain the following expression for the one-loop
effective action
\begin{eqnarray}
&&\Gamma_{div}= -{1\over 2 \epsilon}\int
d^2x\,\sqrt{g}\,\Biggl\{R(N+4)+
                 {2\over
C}V'-{1\over\gamma\Phi}V+\left({f'\over2C}+{f\over4
                 \gamma \Phi } \right)F_{\mu\nu}^{a \, 2} \cr\cr
&&\phantom{\Gamma_{div}=}\quad +N\left({f'\over f}+{1\over \Phi
}-{1\over NC}
                               \right )(\Delta \Phi )       \cr\cr
&&\phantom{\Gamma_{div}=}\quad +N\left({f''\over f}-{{f'^2}\over
f^2}-{1\over
                               \Phi^2}+{C\over
N\gamma\Phi^2}\right)(\nabla^\lambda\Phi )(\nabla_\lambda\Phi)
\Biggr\},
%\label{}
\end{eqnarray}
where $N=\delta_a^a$, and in evaluating the ghost contribution we
have taken into account that $\Gamma_{gh,div}$ is given by the
corresponding expression from Ref. [16], with the only change of
the $R$-term by $\frac{R}{3} (N+8)$ (in the Abelian case $N=1$).

In fact, we have shown that the as long as $Q_\mu^a$ does not mix
with the other fields via gauge conditions, the non-Abelian case
may be obtained
from the
Abelian one by adding quite simple corrections, $\delta\Gamma$,
\begin{equation}
\delta\Gamma= -{N-1\over 2\epsilon}\int
d^2x\,\sqrt{g}\,\Biggl\{ R+ \nabla_\lambda \left[
                         \left({f'\over
f}+{1\over\Phi}\right)(\Delta\Phi) \right]
\Biggr\}.
\label{laste}
\end{equation}
A most remarkable point is that this correction is just a surface
term.

In particular, for the $\Omega$-degenerate De Witt gauge, the
one-loop effective action in 2d dilaton-Yang-Mills gravity is given
by the sum of (60) and of expression (29) with the
$\Omega$-terms excluded. The calculation of the unique effective
action in 2d dilaton-Yang-Mills gravity is more complicated and
will not be done here.

\end{document}